\documentclass[12pt,preprint]{aastex}

\shorttitle{Dynamical mass segregation versus disruption of binary
  stars in dense stellar systems}
\shortauthors{R. de Grijs et al.}

\begin{document}
\title{Gravitational conundrum? Dynamical mass segregation versus
  disruption of binary stars in dense stellar systems}

\author{Richard de Grijs,\altaffilmark{1,2,3} Chengyuan
  Li,\altaffilmark{1,4} Yong Zheng,\altaffilmark{1,4,7} Licai
  Deng,\altaffilmark{5,6} Yi Hu,\altaffilmark{5}\newline
  M. B. N. Kouwenhoven,\altaffilmark{1} and James
  E. Wicker\altaffilmark{5}}

\altaffiltext{1}{Kavli Institute for Astronomy and Astrophysics,
  Peking University, Yi He Yuan Lu 5, Hai Dian District, Beijing
  100871, China; grijs@pku.edu.cn} 
\altaffiltext{2}{Department of Astronomy and Space Science, Kyung Hee
  University, Yongin-shi, Kyungki-do 449-701, Republic of Korea}
\altaffiltext{3}{2012 Selby Fellow, Australian Academy of Science}
\altaffiltext{4}{Department of Astronomy, Peking University, Yi He
  Yuan Lu 5, Hai Dian District, Beijing 100871, China}
\altaffiltext{5}{National Astronomical Observatories, Chinese Academy
  of Sciences, 20A Datun Road, Chaoyang District, Beijing 100012,
  China} 
\altaffiltext{6}{Key Laboratory for Optical Astronomy, National
  Astronomical Observatories, Chinese Academy of Sciences, 20A Datun
  Road, Chaoyang District, Beijing 100012, China}
\altaffiltext{7}{Present address: Department of Astronomy, Columbia
  University, New York, NY 10027, USA}

\begin{abstract}
Upon their formation, dynamically cool (collapsing) star clusters
will, within only a few million years, achieve stellar mass
segregation for stars down to a few solar masses, simply because of
gravitational two-body encounters. Since binary systems are, on
average, more massive than single stars, one would expect them to also
rapidly mass segregate dynamically. Contrary to these expectations and
based on high-resolution {\sl Hubble Space Telescope} observations, we
show that the compact, 15--30 Myr-old Large Magellanic Cloud cluster
NGC 1818 exhibits tantalizing hints at the $\ga 2\sigma$ level of
significance ($> 3\sigma$ if we assume a power-law
secondary-to-primary mass-ratio distribution) of an increasing
fraction of F-star binary systems (with combined masses of 1.3--$1.6
M_\odot$) with increasing distance from the cluster center,
specifically between the inner 10 to 20$''$ (approximately equivalent
to the cluster's core and half-mass radii) and the outer 60 to
80$''$. If confirmed, this will offer support of the theoretically
predicted but thus far unobserved dynamical disruption processes of
the significant population of `soft' binary systems---with relatively
low binding energies compared to the kinetic energy of their stellar
members---in star clusters, which we have access to here by virtue of
the cluster's unique combination of youth and high stellar density.
\end{abstract}


\section{Introduction}

In the absence of gas, gravity is the dominant force driving the
dynamical evolution of stellar systems. Its effects are most easily
discernible in the dense cores of massive stellar clusters. Because of
the close proximity of stars within a cluster, most stars experience
significant gravitational perturbations, close encounters, and
occasionally physical collisions. A cluster's most massive stars are
almost always found in the inner regions (e.g., de Grijs et
al. 2002a,b,c; Gouliermis et al. 2004; and references
therein). However, the origin of this observed mass segregation at
early times and the dynamical timescales required to reach energy
equipartition of at least the most massive stars seem mutually
exclusive, particularly so in the youngest open clusters (Bonnell \&
Davies 1998). These arguments are based on the results of numerical
(`$N$-body') simulations under the assumption of uniform, homogeneous
initial conditions, i.e., taking Plummer spheres as their starting
points. This apparent conflict has led to numerous studies that
explored whether massive stars will most likely form in the centers of
clusters, i.e., through a process coined `competitive accretion,' or
if they might slowly sink to the cluster core owing to gravitational
interactions and energy exchange with other cluster stars, commonly
known as `dynamical mass segregation.' If competitive accretion were
at work, this would possibly require an environmental dependence of
the stellar initial mass function (IMF) on small spatial scales
(referred to as `primordial mass segregation').

Both observations and theoretical arguments suggest that young star
clusters form as highly substructured entities. Observationally,
however, young clusters seem to homogenize on timescales of $<2$ Myr
(Cartwright \& Whitworth 2004; Schmeja et al. 2008). Simulations imply
that this could only happen if clusters are formed {\it dynamically
  cool} (Goodwin et al. 2004; Allison et al. 2009). Several teams
have, therefore, recently performed $N$-body simulations to explore
the earliest phases of cluster evolution (McMillan et al. 2007;
Allison et al. 2009, 2010; Moeckel \& Bonnell 2009a,b; Yu et
al. 2011). They find that initially cool clusters undergo rapid
dynamical mass segregation for stellar masses down to a few solar
masses and within a few million years.

Observations of local areas of active star formation indicate that
almost all stars form in binary or higher-order multiple systems,
across the full stellar mass range (e.g., Kouwenhoven et al. 2005,
2007; Raghavan et al. 2010; Sana \& Evans 2011). These systems are
initially located so close to each other that they interact,
destroying some multiple systems and swapping partners with
others. Such systems could, therefore, significantly affect the
dynamical evolution of a cluster: hard binaries become harder while
soft binaries tend to become softer. The former will have a higher
impact on the dynamical cluster evolution than their equivalent single
stars because of their increased cross section for dynamical
interactions and the combined mass of the binary members. However, the
initial binary fractions, $f_{\rm bin}$, in dense star clusters are
largely unknown; $f_{\rm bin} \equiv B/(S+B+\cdots)$, where $S$ and
$B$ represent single and binary systems, respectively, while the
ellipsis implies inclusion of higher-order multiples. Binary systems
are characterized by mass-ratio distributions $q \equiv m_2/m_1$,
where $m_1$ and $m_2$ are the primary and secondary stellar masses,
respectively, and $m_2 \le m_1$. Since binary systems, and in
particular systems characterized by $q$ relatively close to unity, are
{\it on average} more massive than single stars in a given stellar
population, they are expected to play a more significant role in the
dynamical evolution of their host cluster. In view of the recent
simulations referred to above (Allison et al. 2009; Yu et al. 2011),
one would therefore expect these systems to rapidly mass segregate
dynamically.

\section{Binary systems in young star clusters}

The binary fractions in distant, massive clusters are challenging to
study observationally, although analysis of their color--magnitude
diagrams (CMDs) using artificial-star tests is gaining ground (e.g.,
Zhao \& Bailyn 2005; Sollima et al. 2007; Davis et al. 2008; Hu et
al. 2010; Milone et al. 2012). However, almost all clusters to which
this technique has been applied thus far are old stellar systems in
the Milky Way. Unfortunately, there are no nearby {\it young} massive
clusters, with the possible exceptions of the 4--5 Myr-old massive
cluster Westerlund 1 and the red-supergiant-dominated clusters near
the Galactic Center (e.g., Figer et al. 2006; Davies et al. 2007,
2012). However, all of these young Galactic clusters are affected by
significant foreground extinction and/or forbidding environmental
conditions, so that significant external gravitational effects may
have already altered their stellar make-up, thus preventing us from
assessing the importance of cluster-internal dynamics.

A number of teams have begun to explore the binary fractions in the
young `populous' clusters in the much more distant Large Magellanic
Cloud (LMC; e.g., Elson et al. 1998). We developed and validated an
artificial-star test technique (Hu et al. 2010, 2011) to assess the
binary fractions in dense environments, i.e., by examining a
population's {\it ensemble} properties. Here, we present a detailed
study of the {\it radial dependence} of the fraction of binary systems
characterized by $q \ge 0.55$ in the young (15--30 Myr-old), very
compact (core, effective radii: $R_{\rm core} = 2.1 \pm 0.4$ pc,
$R_{\rm eff} = 5.4$ pc; at the LMC's distance, $1'' \equiv 0.26$ pc),
massive [$M_{\rm cl} = (1.3 - 2.6)\times 10^4 M_\odot$] cluster NGC
1818 using high-resolution {\sl Hubble Space Telescope (HST)}
observations (cf. de Grijs et al. 2002a; Mackey \& Gilmore 2003). In
Hu et al. (2010) we estimated the {\it overall} binary fraction of
F-type stars (1.3--$1.6 M_\odot$) in NGC 1818 at $\sim 0.35$, assuming
a flat mass-ratio distribution for $q > 0.4$, which is consistent with
a total binary fraction of 55 to 100\%.

\section{Hubble Space Telescope observations and analysis}

We used data obtained with the Wide-Field and Planetary Camera-2
(WFPC2) on board the {\sl HST} as part of General Observer (GO)
program GO-7307 (PI Gilmore). WFPC2 contains four chips (each composed
of $800 \times 800$ pixels), a Planetary Camera (PC) and three
Wide-Field (WF) arrays. The PC's field of view is approximately $34
\times 34$ arcsec$^2$ ($0.0455''$ pixel$^{-1}$) and each of the WF
chips has a field of view of approximately $150 \times 150$ arcsec$^2$
($0.097''$ pixel$^{-1}$). We obtained WFPC2 images in the F555W
(henceforth $V$) and F814W ($I$) broad-band filters, with the PC
centered on both the cluster core and on a location offset by
approximately $40''$ to the south west, i.e., roughly pointing at the
cluster's half-mass radius at that location (de Grijs et
al. 2002a). Our exposures in which the PC is located on the cluster
center consist of sets of deep (140 and 300 s for each individual
image in F555W and F814W, respectively) and shallow (5 and 20 s for
each equivalent image, respectively) images. The observations centered
on the cluster's half-mass radius are characterized by individual
exposure times of 800, 800, and 900 s in both filters. The data
reduction has been described by Hu et al. (2010, 2011). Here we use
the same reduced data set.

To explore the radial dependence of the cluster's binary fraction, we
first obtained a new estimate of the cluster center by fitting
Gaussian profiles to the number-density distributions along both the
right ascension ($\alpha_{\rm J2000}$) and declination ($\delta_{\rm
  J2000}$) axes, covering the cluster's area inside $2 R_{\rm eff}$ by
10 bins in each coordinate direction. The resulting center
coordinates, expressed in the WFPC2 world coordinate system, are
$\alpha_{\rm J2000} = 05^{\rm h} 04^{\rm m} 13.2^{\rm s}, \delta_{\rm
  J2000} = –66^\circ 26' 03.7''$.\footnote{Note that this center
  position differs from that given by de Grijs et al. (2002a). This is
  most likely due to the fact that in this paper we report the center
  of the stellar {\it density} distribution while in de Grijs et
  al. (2002a) we determined the center of the {\it luminosity}
  distribution. The latter was based on fitting a two-dimensional
  Gaussian distribution to the heavily smoothed F555W image, assuming
  a symmetric underlying luminosity distribution.} The cluster's
azimuthally averaged number-density profile disappears into the
(approximately constant) background noise at a radius $R = (72.7 \pm
0.3)''$.

Ideally, without binaries and observational errors, all stars in a
cluster that have evolved through the pre-main-sequence phase should
occupy a single isochrone, because they have (approximately) the same
age and metallicity. However, in practice, the cluster's main sequence
is contaminated by field stars. Therefore, we statistically subtracted
background stars (for full details, see Hu et al. 2010). More
importantly, main sequences in observational CMDs tend to exhibit a
non-negligible broadening caused by the presence of true
binary/multiple systems, compounded by a combination of photometric
errors and chance line-of-sight superpositions
(`blending'). Photometric errors cause a symmetrical broadening of the
main sequence, assuming that the magnitude errors are approximately
Gaussian (symmetric with respect to the mean); in our analysis we
adopted an exponential error distribution with increasing magnitude
(Hu et al. 2010). However, chance superpositions and the presence of a
population of physical binary systems both introduce offsets to the
brighter, redder side with respect to the best-fitting
isochrone. Because distinguishing between superpositions and physical
binaries based on CMD morphological analysis alone is all but
impossible, we performed extensive Monte Carlo tests to generate
artificial-star catalogs. We adopted a range of binary fractions (as
well as mass-ratio distributions; see Sect. \ref{massratio.sec}) and
compared the spreads of real and artificial stars with respect to the
best-fitting isochrone.

\begin{figure}
\hspace{2cm}
\includegraphics[width=0.75\columnwidth]{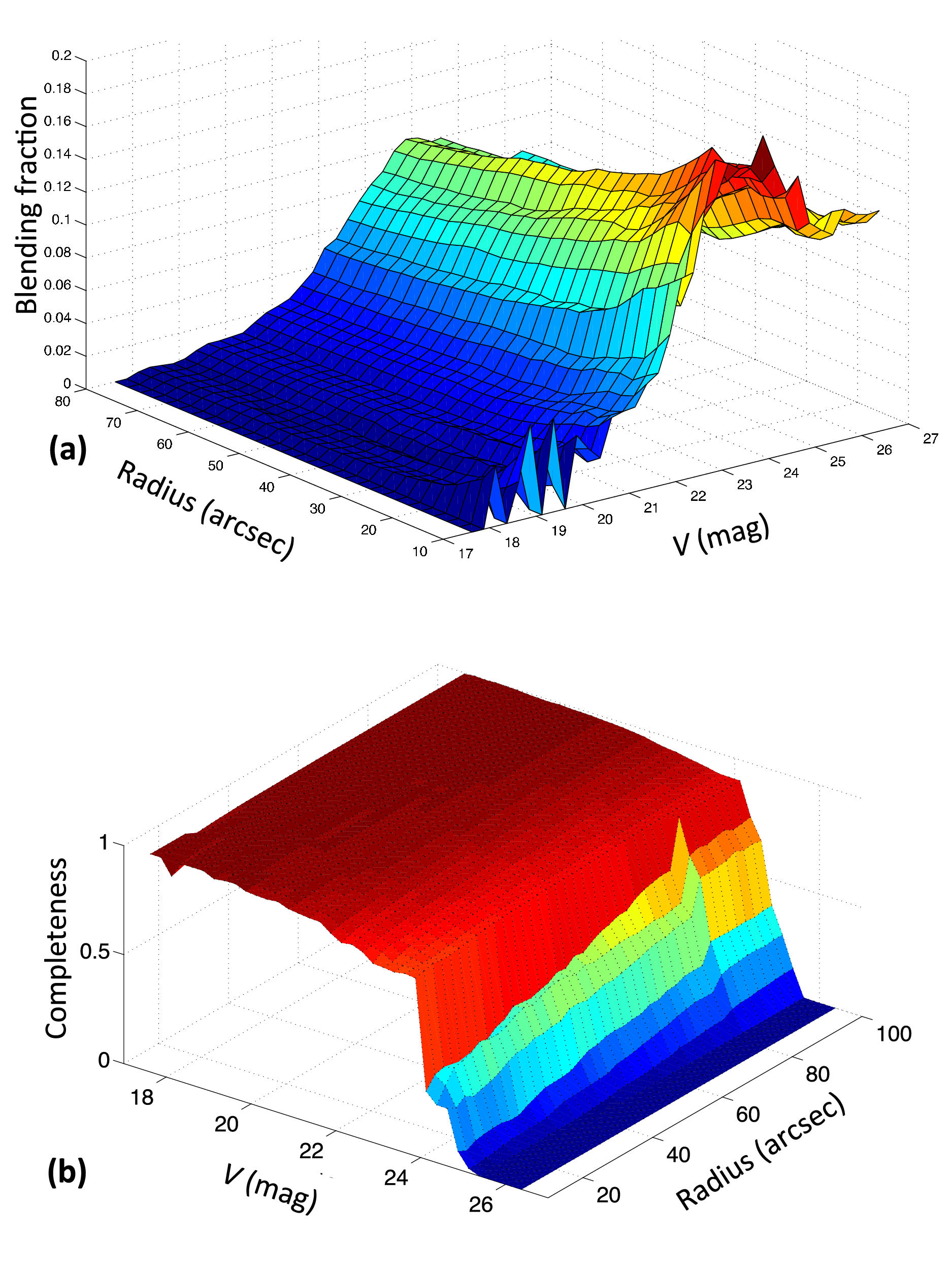}
\caption{\label{blends.fig}(a) Fraction of stellar blends in and (b)
  sample completeness of the NGC 1818 CMD as a function of radius from
  the cluster center and stellar magnitude range.}
\end{figure}

To simulate the effects of chance superpositions and to allow us to
assess the radial dependence of the cluster's binarity, we randomly
distributed a minimum of 640,000 artificial stars, drawn from a Kroupa
(2002)-type stellar IMF covering the stellar mass range from 0.08 to
$50 M_\odot$, on the spatial distribution diagram of the real stars
(for procedural details, see Hu et al. 2010). If the distance between
an artificial star and any real star is $\le 2$ pixels (corresponding
to the minimum separation allowing us to detect them as separate
objects), we assume it to be blended (see Hu et al. 2011). Artificial
stars are not allowed to blend with each other. Fig. \ref{blends.fig}a
shows the resulting blending fraction as a function of both position
within the cluster and the stellar magnitude range sampled. We thus
proceeded to statistically correct the observational NGC 1818 CMD for
the effects of stellar blends. We also determined and corrected our
data for the effects of sample incompleteness as a function of radius
within the cluster (see Fig. \ref{blends.fig}b). Note that the
apparent peak near $V=24$ mag in Fig. \ref{blends.fig}a has been
induced artificially by the effects of incompleteness, as can be
deduced from a comparison with Fig. \ref{blends.fig}b.

In addition, based on extensive tests (see Hu et al. 2010), we found
that any residual background population in the region in CMD space of
interest (see below) is negligible: we derived a systematic error of
$<3$\%. This fraction is based on a comparison of the number of
residual background stars with the number of roughly equal-mass binary
systems in the cluster's CMD.

Star clusters containing equal-mass binary systems will exhibit an
upper envelope to the region in CMD space occupied by binaries which
is 0.752 mag brighter than the locus of the single-star main sequence,
with unequal-mass binaries occupying the space between both
sequences. For our analysis, we selected the region in
color--magnitude space where the single-star main sequence is
shallowest and our sample is $>85$\% complete: for $V < 20$ and $V >
22$ mag, the CMD is too steep to easily disentangle single from binary
stars and blends. In addition, toward fainter magnitudes, photometric
errors start to dominate any potential physical differences, and
field-star contamination becomes increasingly important outside of the
region adopted for our binarity analysis.

\section{Mass-ratio distributions and radial dependences}
\label{massratio.sec}

The remaining free parameter of importance for our analysis is the
mass-ratio distribution, expressed as ${\rm d}N/{\rm d}q \propto
q^{-\alpha}$, where $\alpha > 0 \; (\alpha < 0)$ implies that the
mass-ratio distribution is dominated by low- (high-)mass-ratio binary
systems. Previous studies (Kouwenhoven et al. 2005, 2007; Reggiani \&
Meyer 2011) suggest that $\alpha \in$ [0.0, 0.4] is typical for binary
systems in low-density environments. We will show below that this also
appears a reasonable choice for the binary systems in NGC 1818 and its
surrounding field. Here, we adopt both $\alpha = 0.0$ and 0.4 to
illustrate that the choice of mass-ratio distribution does not
introduce any significant additional uncertainties in our analysis of
the cluster's radial binary fraction. For a given mass-ratio
distribution, the best-fitting binary fraction {\it within a given
  radius} (i.e., a {\it cumulative} binary fraction covering the full
radial range from the cluster center to the radius of interest, a
choice driven by the need to base our results on statistically
significant numbers of stars) is given by statistical $\chi^2$
analysis.

\subsection{Radial trends}
\label{trends.sec}

\begin{figure}
\hspace{1cm}
\includegraphics[width=0.45\columnwidth]{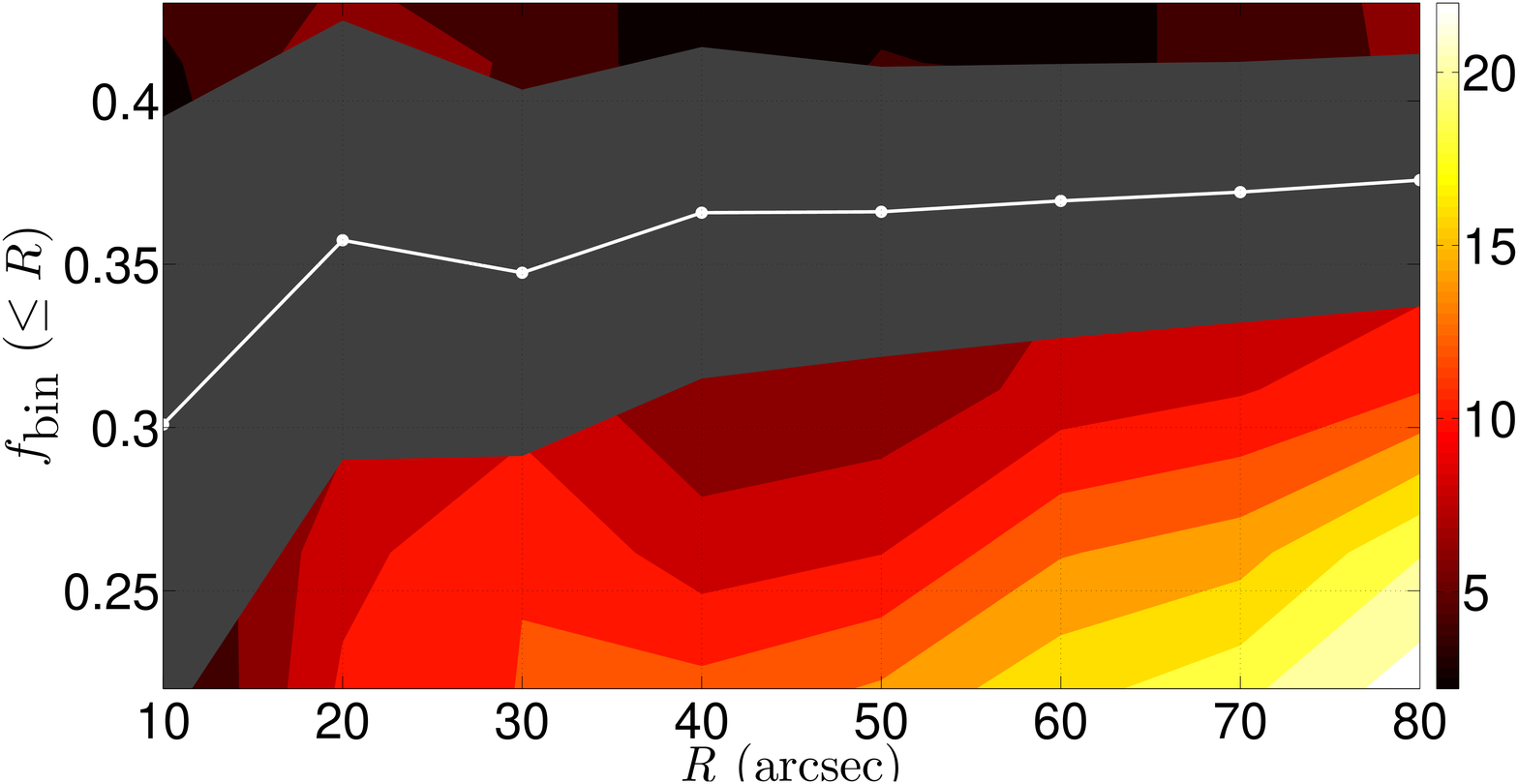}
\hspace{-0.6cm}
\includegraphics[width=0.45\columnwidth]{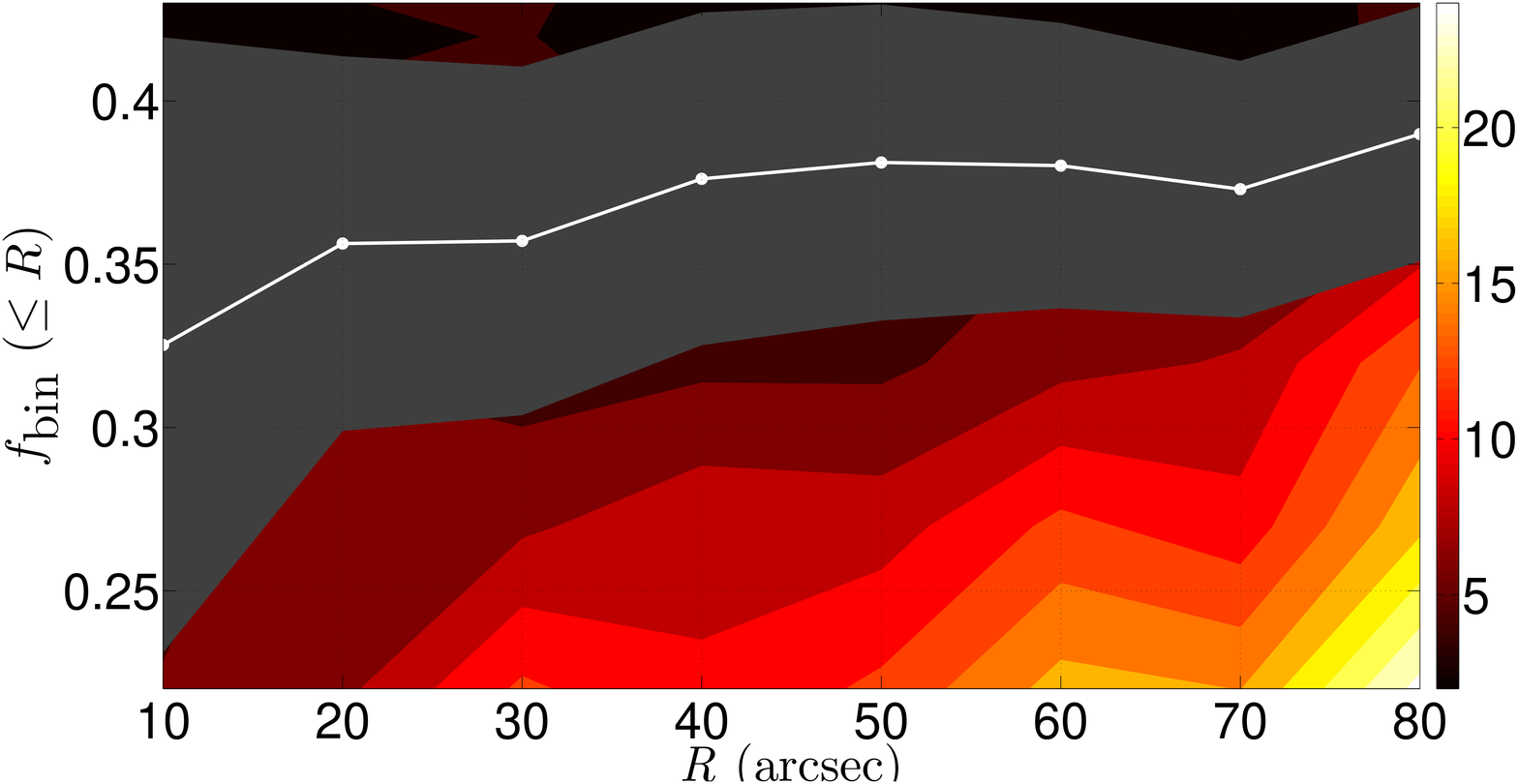}\\
\strut\hspace{1.2cm}
\includegraphics[width=0.8\columnwidth]{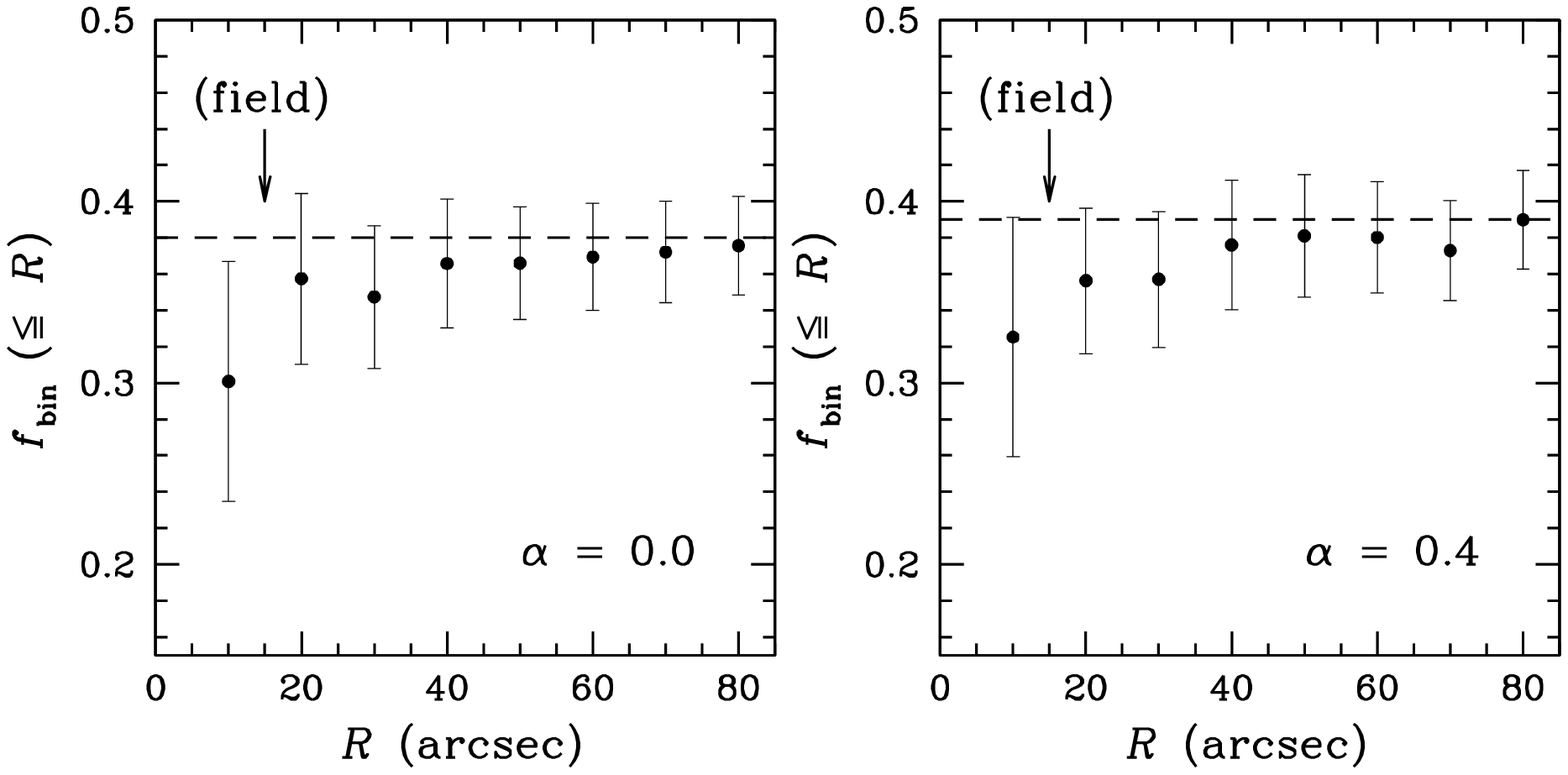}\\
\vspace{-6cm}
\caption{\label{fbin.fig} Cumulative binary fractions in NGC 1818 as a
  function of radius from the cluster center, for power-law indices
  $\alpha = 0.0$ (top left and bottom left) and $\alpha = 0.4$ (top
  right and bottom right), based on $\chi^2$ minimization and for $q
  \ge 0.55$. The color scheme is representative of the $\chi^2$ values
  resulting from our comparison between the observational data and our
  Monte Carlo realizations, with darkest red referring to the lowest
  values. The loci of the minimum $\chi^2$ values and the $1\sigma$
  uncertainties (grey) are also indicated.}
\end{figure}

We generated artificial CMDs based on Padova stellar isochrones
(Bressan et al. 2012) that are representative of the cluster's age and
(roughly solar) metallicity, and characterized by (18) binary
fractions ranging from $f_{\rm bin} = 0.05$ to 0.90 in steps of
0.05. For the data covering the range from the cluster center to a
given radius, we determined the full two-dimensional CMD's $\chi^2$
statistic associated with these variable binary fractions, i.e., we
quantitatively compared the observed and artificial CMDs while only
allowing a single free parameter, $f_{\rm bin}$. For practical
convenience, we used a parabolic curve to describe the dependence of
the $\chi^2$ value of the best fit on the input binary fraction to
obtain both the minimum value, $\chi^2_{\rm min}$, and the $1\sigma$
uncertainties. The latter correspond to the difference between the
binary fractions characterized by $\chi^2_{\rm min}$ and $\chi^2_{\rm
  min}+1$ (Avni 1976; Wall 1996; applicable to single-parameter
fits). The resulting binary fractions as a function of radius in the
cluster, in the restricted magnitude range from $V = 20$ to 22 mag,
and for our two choices of $\alpha$ are shown in Fig. \ref{fbin.fig}
(top: $\chi^2$ landscape for $\alpha = 0.0$ and 0.4; bottom:
corresponding radial binary fractions). Our main result is that the
binary fraction shows hints of an increase out to $40''$, irrespective
of the value of $\alpha \in$ [0.0, 0.4] adopted. Our method is
sensitive to binary systems with $q \ge 0.55$.

Note that the error bars associated with each successive data point as
a function of increasing radius are not statistically
independent. Each new data point includes the stars in our selected
CMD parameter space covered by the data points at smaller radii. The
latter are combined with the stars located at radii beyond those
covered by the previous data point and up to the radius of interest to
yield the cumulative binary fraction at that radius. Specifically, for
$R \le 10''$ and in the magnitude range covered by our Monte Carlo
simulations, the number of stars used for the $\chi^2$ minimization is
92, increasing to 263, 400, 517, 614, 697, 782, and 858 for every
successive cumulative fraction at radii that increase in steps of
$10''$. The $\chi^2$ distribution as a function of (simulated) binary
fraction is very well described by a parabolic function, with a
clearly defined minimum, for all radial ranges.

This, combined with the notion that we are using {\it cumulative}
radial distributions to base our conclusions on, leads us to suggest
that the apparent increasing trend in binary fraction with increasing
radius may be real. At the very least, we can robustly rule out a
decreasing trend as would be expected if dynamical mass segregation
were the main mechanism driving the radial distribution of the stellar
binary fraction, even at (or despite) the cluster's young age. This
would require a conspiracy between the radial distributions of single
and binary systems that is not supported by the observations. Because
of the correlated error bars between successive data points, combined
with the well-defined $\chi^2_{\rm min}$ as a function of $f_{\rm
  bin}(\le R)$, Fig. \ref{fbin.fig} offers tantalizing hints of the
reality of an {\it increasing} fraction of binary systems from the
cluster center outward. If the binary fraction were roughly constant
as a function of radius, in a cumulative distribution such as that
shown here, we would expect to see clearly discernible random changes
in the slope of the distribution between successive data points
instead of the small yet sustained increase potentially detected here.

\subsection{Statistical analysis}

Rather than merely relying on hints, we can make use of
well-established and robust frequentist statistical methods to place
our results on a proper statistical footing. Unfortunately,
application of statistical analysis to the cumulative $f_{\rm bin}$
distributions of Fig. \ref{fbin.fig} is highly complex and, in fact,
poorly understood. Therefore, we proceed by employing individual,
non-overlapping radial ranges for our statistical analysis, so that
the error bars and data points at successive radii are independent.

Nevertheless, our statistical interpretation of the results is not
straightforward, even by making this simplification, because of the
two-step process we employed to obtain the best-fitting $f_{\rm bin}$
values. For every radial range, we independently obtain the $\chi^2$
statistic associated with a set of input $f_{\rm bin}$ values,
employing independently generated random seeds in our Monte Carlo
simulations. For a given $f_{\rm bin}$, the resulting $\chi^2$ value
is thus based on our analysis of a statistically significant number of
stars in color--magnitude space (for the actual numbers of stars used,
see Table \ref{tab1}). In the second step, we fit a parabolic function
to the distribution of $\chi^2$ values as a function of $f_{\rm bin}$
to obtain the mean $\chi^2$ value and its $1\sigma$ uncertainty (as
defined in Section \ref{trends.sec}). The $\chi^2$ distribution at a
given radius is well represented by a symmetric parabolic function.

Our aim is not to ascertain the reality (or otherwise) of a radial
{\it trend}, but to determine whether the means of the $f_{\rm bin}$
values in selected (inner) cluster regions are statistically {\it
  different} from non-overlapping regions elsewhere in the cluster,
{\it given the error bars}. In essence, therefore, for each assessment
we need to compare two samples composed of 18 data points each, i.e.,
the $\chi^2$ values for each of the 18 input $f_{\rm bin}$ values for
a given radial range, corresponding to a total sample size of 36 for
the two distributions being investigated. Since the number of data
points is smaller than approximately 50--60 (the rule-of-thumb lower
limit for adoption of normal distributions; but one should realize
that each of these data points itself was derived on the basis of a
much larger intrinsic sample size), methods based on normal
distributions could prove inadequate because of small-number
statistics. Instead, the Student's $t$ distribution represents the
most appropriate approach to calculate score values associated with
different levels of significance. We compare the resulting $t$ scores
to the relevant values for a $t$ distribution with $(18-1)$ degrees of
freedom. 

The statistical analysis method we have chosen to adopt is a standard
hypothesis-based comparison of two means that follow Gaussian
distributions (e.g., Montgomery 2001, his Chapter 2; Martinez \&
Martinez 2007, their Chapter 6). More sophisticated statistical tests
that analyze the mutual differences at different radii are left for
future work. 

In the current context, the Student's $t$ test is formalized as
follows in a statistically robust manner. Let $\mu_{1}$ and $\mu_{2}$
denote the means of two different minimizations derived from the
$\chi^2$ fits, with standard deviations $\sigma_{1}$ and $\sigma_{2}$,
respectively. We want to connect these expressions to a statistical
value that can be used to test whether or not the means are
statistically equal. In its most general form, the two-sample $t$ test
is well approximated by
\begin{equation}
t_0 = \frac{\mu_{1} - \mu_{2}}{\sqrt{\frac{\hat{\sigma}_{1}^2}{n_{1}}+
    \frac{\hat{\sigma}_{2}^2}{n_{2}}}},
\end{equation}
with 
\begin{equation}
 \nu = \frac{\left(\frac{\hat{\sigma}_{1}^2}{n_1} +
   \frac{\hat{\sigma}_{2}^2}{n_2}\right)^2}
     {\frac{\large(\hat{\sigma}_{1}^2/n_{1}\large)^2}{n_{1} - 1} +
       \frac{\large(\hat{\sigma}_{2}^2/n_{2}\large)^2}{n_{2} - 1}}
\end{equation}
degrees of freedom. In these expressions, $\hat{\sigma}^2$ denotes the
maximum-likelihood estimation of the sample variance, computed by
\begin{equation}
\hat{\sigma}^2 = \frac{\sum_{i=1}^{n} (x_{i} - \mu)}{n-1}.
\end{equation}
To connect the sample variance with the standard deviation derived
from our $\chi^2$ minimization, we assume that we can draw data points
from the distribution derived using $\sigma$. The sample variance of
those data points would be given by $\hat{\sigma}$. Also note that the
denominator of the $t$-test formalism contains an expression of the
form $\hat{\sigma}^2/n$. We can transform this equation as
\begin{equation}
\frac{\hat{\sigma}^2}{n} =
\frac{\sigma^2\hat{\sigma}^2}{\sigma^2 n} =
\frac{\sigma^2\sum_{i=1}^{n} (x_{i} - \mu)}{n \sigma^2 (n-1)} =
\frac{\sigma^2 \chi_{n-1}^2}{n(n-1)}.
\end{equation}
Inserting this transformation into the expression for a $t$ test gives
\begin{equation}
t_0 = \frac{\mu_{1} - \mu_{2}}{\sqrt{\frac{\sigma_{1}^2\chi_{n_{1} -
        1}^2}{n_{1}(n_{1} - 1)} + \frac{\sigma_{2}^2\chi_{n_{2} -
        1}^2}{n_{2}(n_{2} - 1)}}}.
\end{equation}
This expression only contains values that are derived from the
$\chi^2$ minimization. Moreover, we do not need to compute sample data
points, which would be derived from the associated Gaussian
distributions. For the degrees of freedom, we can likewise
substitute the expression for $\hat{\sigma}^2$ to give
\begin{equation}
 \nu = \frac{\left(\frac{\sigma_{1}^2\chi_{n_{1} - 1}^2}{n_1(n_{1} -
     1)} + \frac{\sigma_{2}^2\chi_{n_{2} - 1}^2}{n_2(n_{2} -
     1)}\right)^2} {\frac{\large(\sigma_{1}^2\chi_{n_{1} -
       1}^2/n_{1}\large)^2}{(n_{1} - 1)^3} +
   \frac{\large(\sigma_{2}^2\chi_{n_{2} - 1}^2/n_{2}\large)^2}{(n_{2}
     - 1)^3}}.
\end{equation}
Although this derivation does not constitute a formal proof---a more
rigorous analysis would have to include a full Taylor-series
expansion---this is a convenient approach which gives credence to our
results, since it allows us to connect the number of degrees of
freedom for the $\chi^2$ fits used in our minimization to that used
for the $t$ test we need to use for the statistical analysis. Given
this paradigm, we can then test the hypothesis as to whether or not
the means of the binary fractions are statistically equal at different
radii by applying the $t$-test methodology with associated thresholds.

Under these conditions, let $r_i$ denote radii at various distances
from the cluster center, where $i$ runs from the innermost radial bin
to the bin which encompasses the cluster's outermost radius, here set
to $R = 80''$ for convenience. Let $\mu_i$ and $\sigma_i$ denote the
means and standard deviations at the respective radii. To test if the
mean numbers of binary fractions at different radial ranges exhibit
statistical differences with respect to the outermost bin (denoted by
the subscript `$80''$'), we express a (one-sided) hypothesis test in
the form
\begin{equation}
H_0: \mu_{80''} = \mu_{i} \mbox{ versus } H_1: \mu_{80''} > \mu_{i}.
\end{equation}
The equation that defines values with which the hypothesis can be
tested is
\begin{equation}
t_0 = \frac{\mu_{80''} -
  \mu_{i}}{\sqrt{\frac{\sigma_{80''}^2}{n_{80''}}+
    \frac{\sigma_{i}^2}{n_{i}}}}.
\end{equation}

The $t$ scores are expressed in units of the data set's `quantiles,'
where significance levels of 0.05 and 0.01 correspond to a difference
of 2 or 3 standard deviations between means, respectively. In other
words, significances of 0.10, 0.05, and 0.01 represent confidence
intervals of 90, 95, and 99\%, respectively. Using this methodology,
we compute values for the test statistic and compare these with
threshold values derived from the inverse value of the cumulative
probability-density function of a Student's $t$ distribution. If the
test value is less than the threshold, statistical theory implies that
we must accept the null hypothesis ($H_0$). Otherwise, we are led to
reject the null hypothesis and accept the alternative ($H_1$).

\begin{table}[t!]
\begin{center}
\caption{Student's $t$ scores (expressed in units of the sample's
  quantiles) and levels of significance (`Sign.'; lower limits) for
  radial differences in the mean binary fractions,
  $\mu_i$. Significance levels usually considered statistically
  conclusive (i.e., $> 2 \sigma$) are rendered in {\bf bold-face}
  font. The threshold $t$ scores applicable to a one-sided hypothesis
  test for the 0.10, 0.05, and 0.01 levels of significance are 1.30,
  1.73, and 2.55, respectively.\label{tab1}}
\begin{tabular}{cccccccccc}
\tableline\tableline
\multicolumn{2}{c}{Radial range ($''$)} & &
\multicolumn{2}{c}{$N$(stars)} & & 
\multicolumn{4}{c}{Student's $t$ score} \\
\cline{1-2}\cline{4-5}\cline{7-10}
Inner & Outer & & Inner & Outer & & $\alpha = 0$ & Sign. & $\alpha = 0.4$ & Sign. \\
\tableline
 0--10 & 70--80 & &  92 &  76 & & 2.08 & {\bf 0.05} & 3.25 & {\bf 0.01} \\
 0--10 & 60--80 & &  92 & 161 & & 3.19 & {\bf 0.01} & 3.36 & {\bf 0.01} \\
 0--10 & 40--80 & &  92 & 341 & & 2.37 & {\bf 0.05} & 4.18 & {\bf 0.01} \\
\tableline
 0--20 & 60--80 & & 263 & 161 & & 2.26 & {\bf 0.05} & 1.59 & 0.10 \\
 0--20 & 40--80 & & 263 & 341 & & 0.79 & ---        & 2.26 & {\bf 0.05} \\
\tableline
 0--40 & 40--80 & & 517 & 341 & & 1.55 & 0.10       & 2.53 & {\bf 0.05} \\
\tableline
\end{tabular}
\end{center}
\end{table}

Table \ref{tab1} lists the $t$ scores for the statistical differences
in the mean values between the binary fraction representative of the
cluster's presumably unevolved outer regions (out to $R = 80''$) and
those in other, more central radial ranges considered, for both the
flat mass-ratio distribution and that characterized by a power-law
index $\alpha = 0.4$. For both assumptions of the mass-ratio
distribution, the thresholds for the $t$ scores, assuming a one-sided
hypothesis test (as applied here), are 1.30, 1.73, and 2.55 for the
0.10, 0.05, and 0.01 levels of significance, respectively. We can
interpret these results as follows. Depending on the radial bin size
and to some extent also on the value adopted for the power-law
exponent $\alpha$, it appears that the difference between the mean
binary fractions in the inner 10 to 20$''$ (roughly corresponding to
the cluster's core and half-mass radii, respectively) and the outer 60
to 80$''$ is $\sim 2 \sigma$ (standard deviations), except for the
differences in the means between the inner 10$''$ and any of the outer
radial ranges for the assumption of a power-law mass-ratio
distribution. If the latter assumption holds, the mean binary fraction
in the inner 10$''$ is statistically $\ga 3\sigma$ different from any
of our adopted radial ranges that include the cluster's outermost
radius, $R=80''$. The significance levels for the power-law mass-ratio
distribution are systematically higher than the equivalent values for
a flat mass-ratio distribution. We emphasize that these statistical
results properly account for both the extents of the associated error
bars and the relevant sample sizes.

\subsection{Additional support}

\begin{figure}
\begin{center}
\includegraphics[width=0.5\columnwidth]{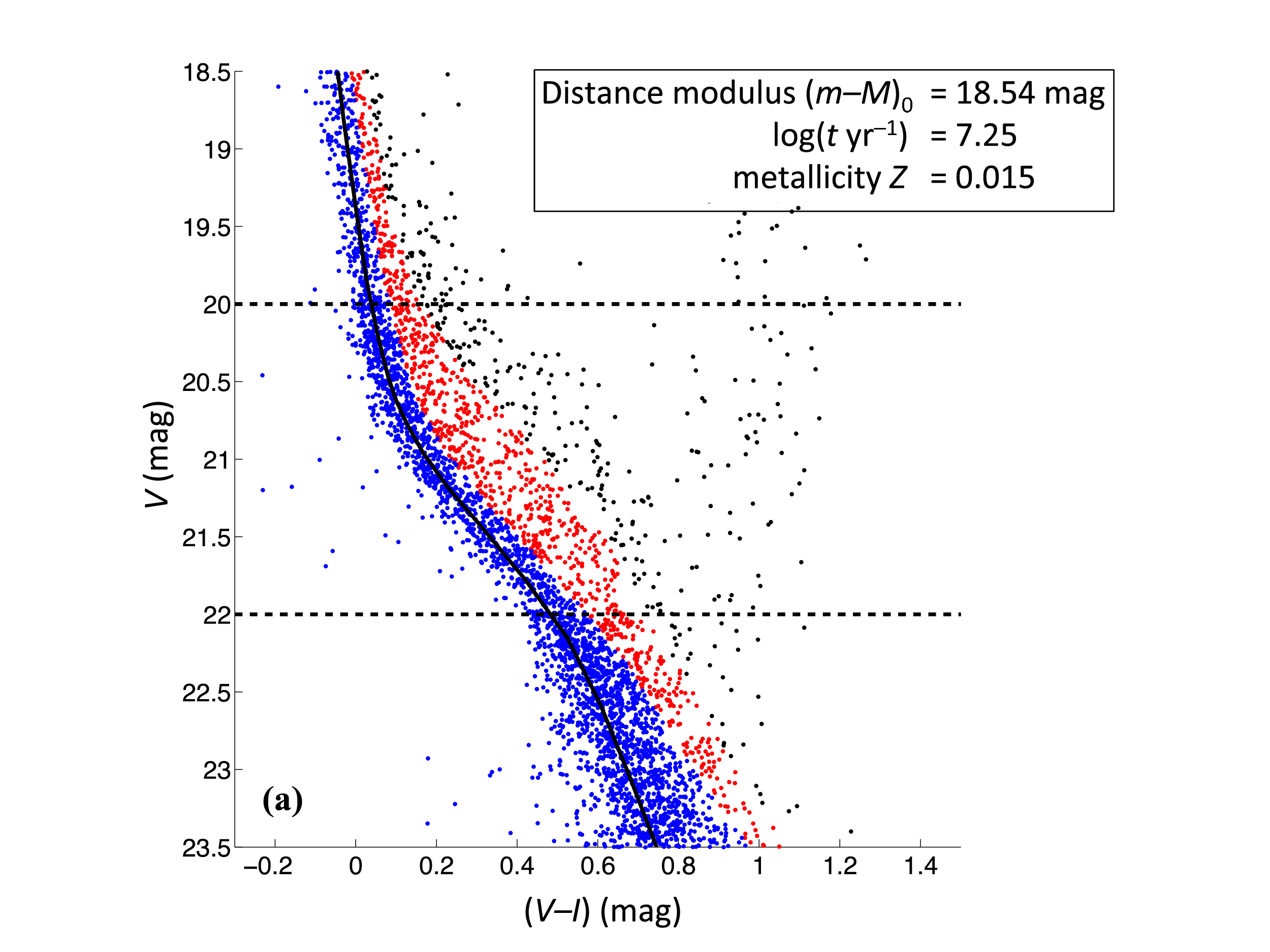}
\includegraphics[width=0.4\columnwidth]{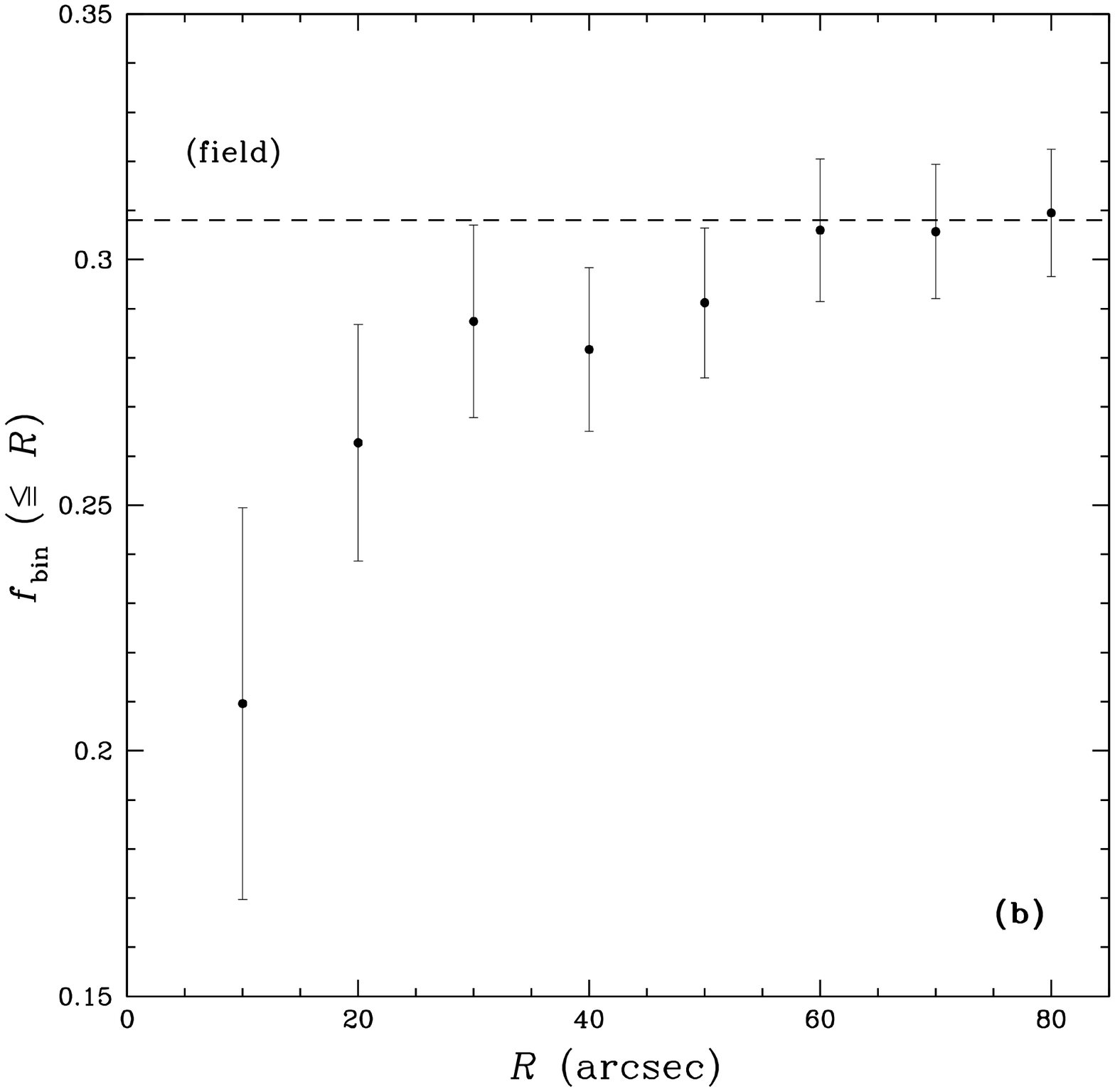}\\
\includegraphics[width=0.7\columnwidth]{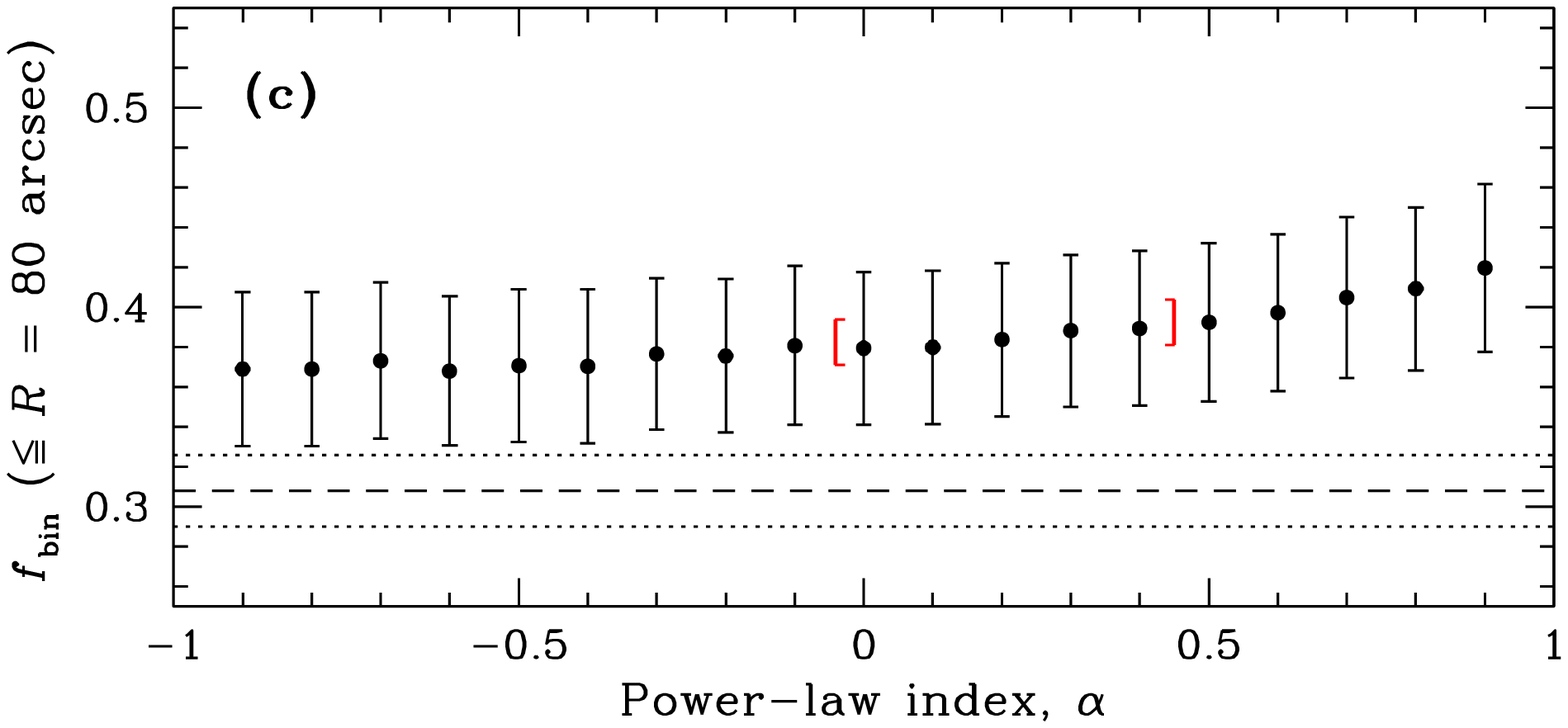}
\end{center}
\vspace{-6cm}
\caption{\label{alpha.fig} (a) NGC 1818 CMD, including the
  best-fitting isochrone (parameters indicated), where we have
  indicated the regions used for our poor-man's approach. Blue:
  single-star area, red: `binaries;' black: discarded stars outside
  our selection regions. (b) Cumulative binary fraction in NGC 1818 as
  a function of radius from the cluster center, based on our
  isochrone-fitting approach. (c) Total binary fraction as a function
  of power-law index, $\alpha$, adopted for the full cluster region
  within $R = 80''$ and for binary systems characterized by $q \ge
  0.55$. The dashed and dotted lines indicate the (lower limit to the)
  binary fraction and the $1\sigma$ error bars, respectively, obtained
  from our isochrone-fitting approach. The (red) square brackets
  indicate the range in $\alpha$ adopted.}
\end{figure}

We subsequently and independently verified the reality of the
suggested rising trend as a function of radius by employing a poor
man's approach to the derivation of the radial binary fractions: we
determined the main-sequence ridgeline and its dispersion,
$\sigma_{\rm ms}$, as well as the expected locus of the equal-mass
binary sequence. We split up the available parameter space into
`single-star' and `binary' regimes. The single-star regime was
delineated by the adopted minimum and maximum magnitudes, and the
main-sequence ridgeline $\pm 3 \sigma_{\rm ms}$ (to account for the
photometric errors). For binary stars, we used the same limiting
magnitudes and adopted the region from the main-sequence ridgeline $+3
\sigma_{\rm ms}$ to the theoretical equal-mass binary sequence $+3
\sigma_{\rm ms}$ (see Fig. \ref{alpha.fig}a). We then proceeded by
counting stars in both areas to obtain a lower limit to the actual
binary fraction, which exhibited a similar increase as a function of
$R$ (see Fig. \ref{alpha.fig}b), flattening out at $31.0 \pm 1.9$\%
for $R \ge 72.7''$, where the cluster profile disappears into the
background noise. 

Although the zero-point calibration of this simple method does not
take into account the effects of blending, the general trend is robust
for $q > 0.6$ (Hu et al. 2010; cf. Elson et al. 1998 for validation of
the $q$ cut). This strengthens the result from our more sophisticated
Monte Carlo approach, so that we conclude that the increasing binary
fraction as a function of radius from the cluster center is indeed
most likely realistic. Note that if this trend were due to incorrect
background corrections, it would follow the cluster's radial density
profile very closely. This is not supported by our results.

Finally, we applied our fitting routines to the full NGC 1818
CMD---within the restricted magnitude range where binary systems are
most clearly discernible and for $R \le 80''$---and examined the
effects of our choice of mass-ratio distribution by varying $\alpha$:
see Fig. \ref{alpha.fig}c. We also included the {\it lower limit}
obtained from our simple isochrone-fitting analysis. This comparison
shows that our choice of $\alpha \in$ [0.0, 0.4] is reasonable; in
this range of $\alpha$, the cluster's derived global binary fraction
is approximately constant within the uncertainties. The binary
fraction obtained from our isochrone-fitting approach returns a lower
limit, because some binary systems inevitably pollute the single-star
regime.
 
\section{Implications}

Numerical simulations based on realistic initial conditions (i.e.,
initial substructure and initially cool dynamics) suggest that
dynamical mass segregation, at least of the most massive stars, is
likely to happen in a crossing time, which is equivalent to the
free-fall time defined by the cluster's gravitational potential. In de
Grijs et al. (2002b), we estimated that NGC 1818 is $\sim$5--30
crossing times old. Hard binary stars may accelerate dynamical mass
segregation significantly, since close encounters between binary
systems and between binaries and single stars are very efficient
(e.g., Parker et al. 2011).

Stars in the stellar mass range targeted in our study, 1.3--$1.6
M_\odot$ (Hu et al. 2010), are not expected to have already reached a
state close to energy equipartition on a cluster-wide scale: the
cluster's half-mass relaxation time for these masses is $\ge 500$ Myr
(de Grijs et al. 2002b). This implies that the process of cluster-wide
dynamical mass segregation is likely still fully underway in NGC
1818. Yet, contrary to dynamical expectations (based on initial
conditions using Plummer spheres), we found a tantalizing hint of an
increasing fraction of binary systems in NGC 1818 from the innermost
(core) radius where we could detect such systems reliably ($\sim
10''$) out to approximately its half-light radius. This is surprising
and, if confirmed independently, flies in the face of previous results
for the same cluster.

A fraction of 35\% ($\pm 5$\%) of roughly similar-mass binaries (with
$17.5 < m_{\rm F555W} < 20.3$ mag, corresponding to $2.0 < m_1 < 5.5
M_\odot$) was reported in the center of NGC 1818 by Elson et
al. (1998), decreasing to 20\% ($\pm 5$\%) beyond $\sim 3 R_{\rm
  core}$. However, our analysis---in particular based on
Fig. \ref{blends.fig}a---has shown that their result can be attributed
to the effects of blending and the near-vertical extent of the stellar
main sequence for their adopted magnitude range, which mask the real
underlying signal. In addition, we already reported a clear detection
of the effects of mass segregation in NGC 1818 for stars with masses
$\ge 1.6 M_\odot$ (de Grijs et al. 2002b), under the assumption that
all stars in our sample were single stars. Despite the (sizeable but
correlated) error bars associated with our main result shown in
Fig. \ref{fbin.fig}, Elson et al.'s (1998) suggested trend cannot be
accommodated by the blending-corrected distribution of single and
binary stars in this cluster.

Since most current theories of binary formation (either dynamically or
primordially) do not explicitly depend on gas or stellar density, we
have no reason to expect {\it a priori} that cluster core environments
feature intrinsically lower binary fractions {\it ab initio}. As such,
we are left to conclude that the suggested trend of an increasing
fraction of F-star binary systems with increasing radius from the
cluster center, if confirmed to be real, may be caused by early
dynamical evolution, i.e., the rapid dissolution of binary systems due
to two-body encounters. The radial dependence of the binary fraction
in dense star clusters has never before been determined for clusters
as young as NGC 1818. We are currently extending our analysis to other
young clusters in the LMC (C. Li et al., in prep.). Preliminary
results for the equivalently young but much more sparsely populated
cluster NGC 1805 indicate a radial trend of greater significance than
for NGC 1818, and which is opposite to the radially increasing
fraction of F-type binaries suggested here (as expected from dynamical
arguments).

If our reported $\ga 2\sigma$ difference in the mean binary fractions
in NGC 1818 between the inner and outer $\sim20''$ stands the test of
further scrutiny, we need to compare the timescale governing dynamical
mass segregation with the expected binary disruption rate so as to
understand this radial dependence. Given the cluster's age in units of
its crossing time (de Grijs et al. 2002b), we expect that its initial
binary population should have been altered by dynamical
interactions. In particular, the destruction of soft (i.e., wide)
binaries due to close encounters---on timescales of order the crossing
time or less (Heggie 1975; Parker et al. 2009)---should be well
underway, at least in the cluster's core region; distant encounters
are unimportant for the disruption of close binaries (Heggie 1975). In
addition, as Heggie (1975) already pointed out, soft binaries are
expected to be more centrally concentrated than single
stars. Therefore, our observed radial dependence suggests that we are
seeing the relatively `hard' binary systems that have survived and may
have been hardened by dynamical encounters. This would offer
unprecedented evidence in support of theoretically predicted dynamical
processes governing star cluster evolution, which we now have access
to by virtue of the unique combination of youth and high stellar
density of NGC 1818.

\section*{Acknowledgements} 
RdG acknowledges useful suggestions from Sungsoo Kim and Eric
Feigelson, while YZ acknowledges fruitful discussions with members of
Jinliang Hou's group at Shanghai Astronomical Observatory. We are
grateful for support from the National Natural Science Foundation of
China through grants 11073001 (RdG, CL), 10973015 (LD), 11003027 (YH)
and 11173004 (MK). MK also received support from the Peter and
Patricia Gruber Foundation (PPGF) through an IAU-PPGF fellowship and
from the Peking University One Hundred Talents Fund (985 program).

Facilities: \facility{HST}

\end{document}